\documentclass[12pt]{iopart} 

\def\ave#1{\langle #1\rangle}
\def\C{{\mathbbm{C}}}
\def\Z{{\mathbbm{Z}}}

\def\P{{\rm P}}
\def\s{{\rm c}}
\def\c{{\rm c}}
\def\e{{\rm e}}
\def\x{{\rm x}}
\def\y{{\rm y}}
\def\z{{\rm z}}
\def\A{{\rm A}}
\def\C{{\rm C}}
\def\S{{\rm S}}
\def\Ns{{\cal N}_\s}
\def\Ne{{\cal N}_\e}
\def\J{{\cal J}}
\def\Js{{\cal J}_\s}
\def\Je{{\cal J}_\e}
\newcommand{\op}[1]{\hat{#1}}

\newcommand{\bra}[1]{\langle #1|}
\newcommand{\ket}[1]{|#1\rangle}
\newcommand{\braket}[2]{\langle #1|#2\rangle}

\newcommand{\tre}{{\, {\rm tr}_{\rm e}\,}} 
\newcommand{\trs}{{\, {\rm tr}_{\rm c}\,}}

\def\ve#1{{\bm{#1}}}

\usepackage{bbm}
\usepackage{bm}
\usepackage{graphicx}

\begin{document}

\title{Decoherence of spin echoes}

\author{Toma\v z Prosen$^1$ and Thomas H. Seligman$^2$}

\address{${}^1$ Physics Department, Faculty of Mathematics and Physics,
University of Ljubljana, Ljubljana, Slovenia\\
${}^2$ Centro de Ciencias Fisicas, University of Mexico (UNAM), Cuernavaca,
Mexico}

\eads{\mailto{prosen@fiz.uni-lj.si}, \mailto{seligman@fis.unam.mx}}  

\begin{abstract}
We define a quantity, the so-called {\em purity fidelity},
which measures the rate of dynamical irreversibility due to decoherence, 
observed e.g in {\em echo} experiments, in the presence of an arbitrary small 
perturbation of the total (system + environment) Hamiltonian. We derive a linear response 
formula for the purity fidelity in terms of integrated time correlation 
functions of the perturbation. Our relation predicts, similarly to the case of
fidelity decay, faster decay of purity fidelity the 
slower decay of time correlations is.
In particular, we find exponential decay in quantum
mixing regime and faster, initially quadratic and later typically 
gaussian decay in the regime of non-ergodic, e.g. integrable 
quantum dynamics. We illustrate our approach by an analytical calculation and numerical
experiments in the Ising spin $1/2$ chain kicked with tilted homogeneous 
magnetic field where part of the chain is interpreted as a system under observation and 
part  as an environment.
\end{abstract}

\submitto{\JPA}

\section{Introduction}

The relation between the rate of decoherence and the nature of dynamics is considered to be quite
established \cite{Zurek1,Zurek2,Zurek3}. In particular, it has been shown \cite{Zurek3} that 
von Neuman entropy of the {\em reduced} (environment averaged) density matrix of an open 
quantum system 
(system + environment) grows (on a short time scale) with the rate given by the Lyapunov exponents
of the corresponding classical dynamics. 

However this is expected to be true only under two rather severe conditions: 
(1) the system has to be initially prepared in a ``coherent'' state (i.e. minimum
uncertainty wave packet) and (2) one has to be deep in the semi-classical regime of very small 
effective Planck constant $\hbar$ since the time-scale of expected quantum-classical correspondence
of the entropy growth, namely the {\em Ehrenfest time}, scales as $t_{\rm E} \propto \log(1/\hbar)$.
Indeed, performing real or even numerical experiments within this regime appears to be very difficult,
see e.g. the numerical experiment in an $N-$atom Jaynes-Cummings model \cite{Nemes}, 
where a difference of the entropy growth between integrable and classically chaotic cases 
has been observed, but its qualitative nature does not become clear. 

On the other hand, by rejecting the assumption (1) on the coherent initial state  
but considering a random pure initial state instead, one can use orthogonal/unitary invariance and  
define simple random matrix models to analyze the time evolution of decoherence \cite{GS}, 
and no qualitative and minimal quantitative 
differences between regular and chaotic dynamics were observed. 
Furthermore, we should stress that in quantum information science considering random initial states  
is potentially much more useful. Indeed, in order to use a massive parallelism of quantum computation 
one has to prepare the quantum computer system initially in a coherent superposition (pure state)  
of as many as possible elementary qubit (basis) states, but to contain maximal information 
this state can and should be any possible state and thus will behave more 
like a random state than like  a gaussian wave packet. 
 
Based on a proposition by Peres \cite{Peres2}, there has been much interest in viewing  
reversibility of a process with an imperfection on the reversed time evolution rather than on the  
state as discussed by Casati {\em et al} \cite{casati}.  
This allows to avoid some of the essential implications of linearity of quantum mechanics 
that trivialize the latter case. The related {\em state correlation function} is usually called 
{\em fidelity}.
Its behaviour for chaotic systems on the Ehrenfest time scale has first 
been discussed in \cite{Jalabert} with similar findings as in \cite{Zurek1,Zurek2,Zurek3}. 
On the other hand it has been found recently \cite{Prosen01,QC,PZ} that on more relevant 
time scales related to the decay of correlation functions of quantum observables in the 
mixing case the fidelity decay is indeed exponential, but with 
a very different exponent, which is determined basically by the strength of the perturbation. 
This result was obtained for any state and, what is more, for the 
integrable case a faster gaussian decay was found in the thermodynamic limit\cite{Prosen01}. 
More strictly speaking the result implied for {\em small times} or {\em small perturbation strengths}
linear decay behaviour in the mixing case and 
quadratic decay in the integrable or more generally non-ergodic case, where `small times' may be 
still large compared to the perturbative regime where all dependencies are quadratic anyway.

There is a widespread sense that decoherence should follow correlation functions \cite{Schleich}, 
and we therefore expect that in the present situation decoherence 
should behave along the same lines as fidelity. Yet it is worthwhile to test this, as understanding 
decoherence is essential for the quantum information applications. 
The central issue of this paper is 
thus to test the evolution of decoherence of echoes (some aspects of which are relevant in the 
real spin-echo experiments \cite{Usaj}). Indeed it is crucial to determine if  
fidelity is a reliable measure. This is all the more true because of the counter-intuitive 
result that fidelity is higher for mixing systems than for integrable ones.  
Due to the great interest in quantum information problems  
we shall mainly focus on random states for the reason pointed out above.  
 
Basically we follow previous work on fidelity \cite{Prosen01,QC,PZ} 
and  express our results on decoherence by time correlation  
functions of quantum observables, 
though we shall have to generalize the concept of correlation 
function slightly. We shall use the  
purity \cite{Zurek2} (which is one minus the linear entropy or idempotency defect  
used in  \cite{Nemes}) instead of the von Neumann 
entropy as a measure of decoherence, 
because purity, being an analytic function of the density matrix, is much easier to handle. 
Rather than introducing non-unitary time-evolution we follow the simpler way used in 
\cite{Nemes,GS} of partial tracing over the environment  
in a product Hilbert space after unitary time evolution. 
The kicked spin chain will serve as a specific example for our 
considerations, where analytic calculations for the integrable case can be 
carried further and numerics will confirm the validity of our 
approximations. 
 
The unitary propagator $U$ of the total system (central system + 
environment) can be either a short time-propagator 
$U = \exp(-i H \Delta t/\hbar)$, or a Floquet map  
$U = \op{\cal T}\exp(-i\int_0^p d\tau H(\tau)/\hbar)$ for a  
periodically time-dependent Hamiltonian $H$ ($H(\tau+p)=H(\tau)$), or any abstract quantum map. 
We only assume that the total Hilbert space ${\cal H}$ (the domain of $U$) can be written 
as a direct product of two parts 
\begin{equation} 
{\cal H} = {\cal H}_\s \otimes {\cal H}_\e. 
\end{equation} 
In what follows, the subscripts `c' and `e' will denote quantities referring to  
the {\em central system} and 
the {\em environment}, respectively, and to the total system when no subscript will be attached. 
 
As mentioned above we shall not introduce a non--unitary time evolution, but just 
consider the decoherence induced in the central system by its entanglement with the environment 
and then perform partial traces over the latter. In other words we shall 
test the stability of the  
disentanglement caused by inverse time-evolution operator if this operator is perturbed. 
This perturbation can generally be described by some self-adjoint operator $A$  
over ${\cal H}$ 
and some perturbation strength $\delta$ to yield
\begin{equation}
U_\delta = U\exp(-i A\delta/\hbar).
\label{eq:U_d}
\end{equation}     
Our perturbation is static (the generator $A$ has no explicit time-dependence) and cannot  
be associated to noise  
since it is not due to the coupling to the environment (which is explicitly included from the  
outset) but stems  
from the imperfect (or unknown) description of the total hamiltonian. 
 
First, let us consider coherent quantum evolution of the total system, and  
define the {\em fidelity} as 
an overlap between states evolving under unperturbed and perturbed 
time-evolution  
\begin{equation} 
F(t) = \braket{\psi_\delta(t)}{\psi(t)} = \bra{\psi} U^{-t}_\delta U^t \ket{\psi} 
\label{eq:fid} 
\end{equation} 
where our time $t$ is an integer. 
Fidelity has become a standard measure of instability of quantum computation \cite{qcomp},  
but has also been 
considered in a more abstract context as a measure for the instability of quantum dynamics 
\cite{Peres2,Jalabert,Pastawski,Prosen01,QC,Beenakker,Tomsovic,PZ,cucchietti,cohen,benenti}. 
It has recently been pointed out \cite{Prosen01,QC,PZ} that fidelity is  
intimately related to the decay of correlations, and that, surprisingly enough, fidelity  
decay is faster the  
slower decay of correlations is. However, fidelity is a property of a pure state of the  
total system  
which is typically not accessible. What can defacto be measured is only information relating 
to the reduced density operator 
\begin{equation} 
\rho_\c(t) = \tre \ket{\psi(t)}\bra{\psi(t)} 
\label{eq:rhot} 
\end{equation} 
where $\tre$ is a {\em partial trace} over the environmental degrees of freedom ${\cal H}_\e$. 
 
We now have to obtain an extension of the concept of fidelity in order to measure coherence  
properties of the reduced density matrix.  Assume  we prepare our system in a pure product 
(disentangled) state  
\begin{equation} 
\ket{\psi}=\ket{\psi_\c}\otimes\ket{\psi_\e},  
\label{eq:inistate} 
\end{equation} 
such that the reduced density matrix is also pure $\rho_\c(0) = \ket{\psi_\s}\bra{\psi_\s}$. Then  
we propagate our  
system for some time $t$, after which the state $U^t\ket{\psi}$ becomes generally an 
entangled superposition of  
system and environment states. We then invert time, {\it i.e.} we change the sign of the  
Hamiltonian with a small inaccuracy described by the operator $A$ 
(\ref{eq:U_d}), and propagate the  
system backwards for the same amount of time arriving at the final `echo' state  
\begin{equation} 
\ket{\phi(t)} = U^{-t}_\delta U^t\ket{\psi}. 
\end{equation}  
The overlap between the initial and the final state would be just a fidelity (\ref{eq:fid}),  
$F(t) = \braket{\phi(t)}{\psi}$, however here we are interested only to what 
extent has the final state disentangled from the environment {\it i.e.} how 
much the final reduced density operator 
\begin{equation} 
\rho_\c^{\rm echo}(t) = \tre \ket{\phi(t)}\bra{\phi(t)}. 
\end{equation} 
deviates from a pure state. This is best quantified in terms of a {\em purity},  
$ \tr\rho^2$, which is equal to $1$ for a pure state 
and less than $1$ otherwise. The minimal value of purity is $1/\Ns$ where $\Ns$ is the  
 dimension of ${\cal H}_\c$; note that this limiting value can only be reached if the  
dimension of the environment $\Ne$
tends to infinity. For finite dimensional environment space we find later 
a more accurate limiting value (\ref{eq:Fstar}) (see also \cite{GS}).
We shall therefore  study, {\em the purity fidelity} defined  as the purity of an  
echoed reduced state 
\begin{equation} 
F_\P(t) = \trs \left[\rho_\c^{\rm echo}(t)\right]^2 = 
\tr_\s \left[\tr_\e \left(U_\delta^{-t} U^{t} \ket{\psi_\c}  
\ket{\psi_\e} \bra{\psi_\c}\bra{\psi_\e}U^{-t} U^{t}_\delta\right)\right]^2  
\label{eq:purfid} 
\end{equation} 
 
In section 2 we shall make our main theoretical predictions on purity fidelity in relation  
to ergodic properties of  
dynamics and in particular to the correlation decay. In section 3 we shall apply our results  
in the quantum spin chain model, namely 
the Ising spin $1/2$ chain kicked with a tilted homogeneous magnetic field, which exhibits all  
qualitatively different 
regimes of quantum dynamics ranging from integrable to mixing.  
In section 4 we shall discuss some implications of our results and conclude. 
 
\section{The relation between purity fidelity and correlation decay} 
 
\subsection{Linear response} 
 
Following \cite{Prosen01,QC,PZ} we start from time-dependent perturbation theory 
(linear response) and expand the purity fidelity in 
a power series in the perturbation strength.  
It is convenient to write the complete basis of ${\cal H}$ as $\ket{j,\nu}$, where Latin/Greek  
indices running over 
${\cal N}_{\c,\e} = \dim{\cal H}_{\c,\e}$ values denote system/environmental degrees of freedom,  
such that the initial 
state is always designated as $\ket{1,1} = \ket{\psi_\c}\otimes\ket{\psi_\e}$. 
First, let us rewrite the purity fidelity 
\begin{eqnarray} 
F_\P(t) &=& \tr_\s \left[\tr_\e \left(M_t \ket{\psi_\s}\ket{\psi_\e}\bra{\psi_\c}\bra{\psi_\e} M_t^\dagger\right)\right]^2 = 
\label{eq:purfidm} \\ 
        &=& \sum_{j,k,\mu,\nu}\bra{j,\mu}M_t\ket{1,1}\bra{1,1}M^\dagger_t\ket{k,\mu}\bra{k,\nu}M_t\ket{1,1} 
\bra{1,1}M_t^\dagger\ket{j,\nu}\label{eq:purfid2} 
\end{eqnarray} 
in terms of a {\em unitary fidelity operator} 
\begin{equation} 
M_t = U^{-t}_\delta U^t. 
\label{eq:m} 
\end{equation} 
Second, we observe \cite{Prosen01} that the fidelity operator can be rewritten in terms  
of the perturbing operator in  
the Heisenberg picture $A_t = U^{-t} A U^t$, namely 
\begin{equation} 
M_t = \exp(i A_0\delta/\hbar)\exp(i A_1\delta/\hbar)\cdots\exp(i A_{t-1}\delta/\hbar).  
\label{eq:Mprod} 
\end{equation} 
This expression can be formally expanded into a power series 
\begin{equation} 
M_t = 1 + \sum_{m=1}^\infty \frac{i^m \delta^m}{m!\hbar^m} 
 \op{\cal T} \!\!\! \sum_{t_1,\ldots,t_m=1}^{t} A_{t_1} A_{t_2} \cdots A_{t_m}, 
\label{eq:Msum} 
\end{equation} 
which always converges (in the {\em strong limit} sense) provided the generator $A$ is a bounded operator. 
The symbol $\op{\cal T}$ denotes a left-to-right time ordering of operator products. 
Third, we truncate the expression (\ref{eq:Msum}) at the second order $\delta^2$ and plug  
it into expression  
(\ref{eq:purfidm}). After a tedious but straightforward calculation we find 
\begin{equation} 
F_\P(t) =  1 - \frac{2\delta^2}{\hbar^2}\sum_{t',t''=0}^{t-1}\sum_j^{j\neq 1}\sum_\nu^{\nu\neq 1} 
\bra{1,1}A_{t'}\ket{j,\nu}\bra{j,\nu}A_{t''}\ket{1,1} + {\cal O}(\delta^4). 
\label{eq:lr1} 
\end{equation} 
The RHS of this linear response formula indeed looks much like a time correlation 
function, however with 
funny exclusion rules on the state summation. It can be written more elegantly 
in a basis independent way as 
\begin{equation} 
F_\P(t) = 1 - \frac{2\delta^2}{\hbar^2} \tr\left[\rho\,\Sigma A_t\,\tilde{\rho}\,\Sigma A_t\right] +  
{\cal O}(\delta^4), 
\label{eq:lr2} 
\end{equation} 
where $\Sigma A_t\equiv\sum_{t'=0}^{t-1} A_{t'}$, $\rho=\ket{\psi}\bra{\psi}$, and 
$\tilde{\rho}=(\mathbbm{1}_\s-\ket{\psi_\s}\bra{\psi_\s})\otimes 
              (\mathbbm{1}_\e-\ket{\psi_\e}\bra{\psi_\e})$. 
Our expectation is that the properties of purity fidelity decay will mainly depend on 
dynamics, i.e. the behavior of the operator $\Sigma A_t$, and less
on the detailed structure of the initial state encoded in $\rho$ and $\tilde{\rho}$. 
Let us first discuss the limiting qualitatively different cases of dynamics: 
 
\subsubsection{Regime of ergodicity and mixing.}  
In the regime of {\em ergodic and mixing} quantum dynamics \cite{qmix}, the  
{\em reduced transport coefficient}  
\begin{equation} 
\tilde{\sigma} = \lim_{t\to \infty}\frac{1}{2t}\tr[\rho\,\Sigma A_t\,\tilde{\rho}\,\Sigma A_t] 
\label{eq:sigma} 
\end{equation} 
can be estimated in terms of the usual Kubo transport coefficient 
\begin{equation} 
\sigma = \lim_{t\to \infty}\frac{1}{2t}\sum_{t',t''=0}^{t-1}\tr[\rho A_{t'} A_{t''}] 
\end{equation} 
which is obviously finite $\sigma < \infty$ in the case of mixing dynamics and 
sufficiently strong decay of time correlations. 
Namely one can easily prove that $0 \le \tilde{\sigma} \le \sigma$ 
since all the terms of $\sigma$ missing in $\tilde{\sigma}$, as well as all the terms of 
$\tilde{\sigma}$ when expanding along (\ref{eq:lr1}), are non-negative. 
For times $t$ larger than a certain {\em mixing time} $t_{\rm mix}$, $t > t_{\rm mix}$,  
i.e. a characteristic time scale on which the limiting process (\ref{eq:sigma}) converges, 
we thus find a linear decay of purity fidelity 
\begin{equation} 
F_\P(t) = 1 - \frac{4\delta^2}{\hbar^2}\tilde{\sigma} t + {\cal O}(\delta^4) =  
1 - \frac{4t}{\tilde{\tau}_{\rm em}} + {\cal O}(\delta^4),  
\label{eq:lrem} 
\end{equation} 
on a time scale  
\begin{equation} 
\tilde{\tau}_{\rm em} = \frac{\hbar^2}{\delta^2\tilde{\sigma}} \propto \delta^{-2}. 
\end{equation} 
 
\subsubsection{Non-ergodic regime.} In the opposite regime of {\em non-ergodic}  
(e.g. integrable) quantum dynamics, 
the non-trivial (e.g. different than a multiple of identity) time-averaged operator exists 
\begin{equation} 
\bar{A} = \lim_{t\to\infty} \frac{1}{t}\Sigma A_t, 
\label{eq:tave} 
\end{equation} 
so that for times larger than a certain {\em averaging time} $t > t_{\rm ave}$,  
i.e. a characteristic time scale on which the limiting process (\ref{eq:tave}) converges, 
we find a quadratic decay of purity fidelity 
\begin{equation} 
F_\P(t) = 1 - \frac{2\delta^2}{\hbar^2} \tilde{D} t^2 + {\cal O}(\delta^4) =  
1 - 2\left(\frac{t}{\tilde{\tau}_{\rm ne}}\right)^2 
+ {\cal O}(\delta^4), 
\label{eq:lrne} 
\end{equation} 
with a coefficient which we name as {\em reduced stiffness}
\begin{equation}
\tilde{D} = \tr[\rho\bar{A}\tilde{\rho}\bar{A}]\;\; 
\end{equation}
on a time scale  
\begin{equation} 
\tilde{\tau}_{\rm ne} = \frac{\hbar}{\delta \sqrt{\tilde{D}}} \propto \delta^{-1}.  
\end{equation} 
We note that our reduced stiffness coefficient is again rigorously bounded by 
the common (non-reduced) stiffness $D=\tr[\rho\bar{A}^2]$, 
namely $\tilde{D} \le D$.
 
\subsection{Beyond the linear response} 
 
Next we extend our linear-response results to the regime, where the value of purity fidelity  
becomes considerably lower than $1$. Again we consider the two qualitatively different 
cases of dynamics:

\subsubsection{Regime of ergodicity and mixing.} Here we assume, in full analogy with
the derivation of fidelity decay \cite{Prosen01,PZ}, that also the property of quantum 
$n$-mixing holds (in the asymptotics ${\cal N}\to\infty$), namely that all the higher 
$n-$point time correlation function factorize so that the $\delta-$expansion of the
fidelity operator can be summed up and in the {\em weak limit} sense we
obtain
\begin{equation} 
e^{t/\tau_{\rm em}} M_t \to \mathbbm{1}, \quad {\rm as}\quad t\gg t_{\rm mix},  
\quad {\rm where}\quad 
\tau_{\rm em} = \frac{\hbar^2}{\delta^2\sigma} 
\label{eq:Mtem} 
\end{equation} 
This statement is equivalent to the statement shown in \cite{Prosen01,PZ} that for  
arbitrary (pure or mixed) state $\rho$, 
fidelity decay $F(t\gg t_{\rm mix}) = \tr \rho M_t = \exp(-t/\tau_{\rm em})$ is independent 
of the state. 
In case of a semi-classical situation of small effective value of $\hbar$ the limit  
(\ref{eq:Mtem}) starts to build up \cite{PZ} only after the Ehrenfest time 
$t_{\rm E}$, $t_{\rm E} = \ln(1/\hbar)/\lambda$, where $\lambda$ is the 
largest classical Lyapunov exponent, so then $t_{\rm mix} = t_{\rm E}$. 
Of course, (\ref{eq:Mtem}) never holds in 
the strong limit sense as the fidelity operator $M_t$ is unitary. 
Plugging (\ref{eq:Mtem}) into formula (\ref{eq:purfidm}) we obtain for the  
asymptotic exponential decay of purity fidelity 
\begin{equation} 
F_\P(t) = \exp(-4t/\tau_{\rm em}), \quad {\rm as}\quad t\gg t_{\rm mix}. 
\label{eq:FPem} 
\end{equation} 
Comparing eqs. (\ref{eq:FPem}) and (\ref{eq:lrem}) in the asymptotic regime  
$t\to\infty$ while approaching $\delta\to 0$ so as 
to remain in the regime of linear response we find that we must have strictly  
$\tilde{\tau}_{\rm em}=\tau_{\rm em}$, i.e 
\begin{equation} 
\tilde{\sigma}=\sigma. 
\label{eq:eqsig} 
\end{equation} 
Of course, this argument is correct only for sufficiently strong mixing, 
{\it e. g.} for exponential decay of 
correlations. 
 
\subsubsection{Non-ergodic regime.} Again, in analogy to the simple fidelity decay in  
non-ergodic situations\cite {Prosen01,PZ}, we can rewrite the fidelity operator  
in terms of the time averaged operator  
$\bar{A}$, 
\begin{equation} 
M_t \to \exp(i\bar{A}t\delta/\hbar),\quad {\rm as}\quad t\gg t_{\rm ave}. 
\label{eq:Mtne} 
\end{equation} 
Plugging the fidelity operator (\ref{eq:Mtne}) 
into formula (\ref{eq:purfidm}) we find that, as a manifestation of
non-ergodicity, the behaviour of purity fidelity generally depends on the structure of the initial state.
 
\subsection{State averaged purity fidelity} 

As we stated in the introduction, we are interested in the behaviour of a 
random initial state of the central system and as far as the environment is 
concerned we as well average over states to reflect our ignorance of the 
latter.
Let us therefore start with a {\em random product initial state}  
(\ref{eq:inistate}) implying that the 
states $\ket{\psi_\c}$ and $\ket{\psi_\e}$ are random ${\cal N}_\c$ and  
${\cal N}_\e$ dimensional vectors, whose 
components $\braket{j}{\psi_\c}$ and $\braket{\mu}{\psi_\e}$ are, in  
the limits ${\cal N}_\c\to\infty$ and 
${\cal N}_\e\to\infty$, {\em independent complex random gaussian variables} with 
variance $1/\Ns$ and $1/\Ne$, respectively. 
Denoting averaging over random product initial state as 
$\ave{.}_\psi$ one can easily average the linear response formula (\ref{eq:lr1}) 
\begin{equation} 
\ave{F_\P(t)}_\psi = 1 - \frac{2\delta^2 \Ns \Ne}{\hbar^2 (\Ns+1)(\Ne+1)}\sum_{t',t''=0}^{t-1} C_\P(t',t'') 
+ {\cal O}(\delta^4) 
\label{eq:lra} 
\end{equation} 
expressing the purity fidelity decay in terms of a sum of {\em reduced correlation function} 
\begin{eqnarray} 
C_\P(t',t'') &=& \ave{A_{t'} A_{t''}} + \ave{A}^2 -  
\ave{\ave{A_{t'}}_\e \ave{A_{t''}}_\e}_\c - \ave{\ave{A_{t'}}_\c \ave{A_{t''}}_\c}_\e, \label{eq:CP} 
\end{eqnarray} 
where $\ave{.}_\c \equiv (1/\Ns)\trs(.)$, $\ave{.}_\e \equiv (1/\Ne)\tre(.)$, and  
$\ave{.} = \ave{\ave{.}_\e}_\c =  
(1/{\cal N})\tr(.)$ 
 
Again, in ergodic and mixing regime, where correlation functions (\ref{eq:CP})  
decay fast as $|t''-t'|\to\infty$, 
the average purity fidelity exhibits initial linear decay (\ref{eq:lrem}) with average 
transport coefficient 
\begin{equation} 
\ave{\tilde{\sigma}}_\psi = \lim_{t\to\infty}\frac{1}{2t}\sum_{t',t''=0}^{t-1}C_\P(t',t''). 
\end{equation} 
Since, for sufficiently strong mixing we have (\ref{eq:eqsig}), 
$\ave{\sigma}_\psi = \lim_{t\to\infty}\frac{1}{2t}\sum_{t',t''=0}^{t-1}C(t'-t'') = \ave{\tilde{\sigma}}_\psi$, 
where $C(t) = \ave{A_t A} - \ave{A}^2$, then  
$\lim_{t\to\infty}\frac{1}{t}\sum_{t',t''=0}^{t-1}(C(t'-t'')-C_\P(t',t''))=0$. 
We note again that all the results referring to the ergodic and mixing regime implicitly 
assume the limits $\Ns,\Ne\to\infty$ to be considered before $t\to\infty$.
 
On the other hand, in non-ergodic regime we have quadratic initial decay (\ref{eq:lrne}) with state-averaged
reduced stiffness
\begin{equation}
\ave{\tilde{D}}_\psi = \frac{\Ns \Ne}{(\Ns+1)(\Ne+1)}
\left(\ave{\bar{A}^2} + \ave{\bar{A}}^2 - \ave{\ave{\bar{A}}_\c^2}_\e - \ave{\ave{\bar{A}}_\e^2}_\c\right).
\label{eq:redstif}
\end{equation}

Without making reference to any of the two extreme cases the general expression
for the purity fidelity (\ref{eq:purfid2}) may be simply 
state averaged while keeping the fidelity operator completely general:
\begin{eqnarray}
\ave{F_\P(t)}_\psi &=& \frac{\Ns + \Ne}{(\Ns + 1)(\Ne + 1)} +
\frac{1}{\Ne(\Ne+1)\Ns(\Ns+1)}\times \label{eq:FPSA} \\
\sum_{j,k,p,q,\alpha,\beta,\mu,\nu}&\Bigl(&\!\!
\bra{p,\alpha}M_t\ket{j,\nu}\bra{j,\mu}M_t^\dagger\ket{q,\alpha}\bra{q,\beta}
M_t\ket{k,\mu}\bra{k,\nu}M_t^\dagger\ket{p,\beta}
\nonumber\\&+&\!\!
\bra{p,\alpha}M_t\ket{j,\nu}\bra{k,\nu}M_t^\dagger\ket{q,\alpha}\bra{q,\beta}
M_t\ket{k,\mu}\bra{j,\mu}M_t^\dagger\ket{p,\beta}
\Bigr)
\nonumber
\end{eqnarray}
From (\ref{eq:FPSA}) one observes that (averaged) purity fidelity does not decay 
to zero. It has a
finite saturation plateau value $F^*$, which can be estimated by assuming 
that after a long time $t$ and for sufficiently 
strong perturbation $\delta$ (i.e. such that the eigenstates of $U_\delta$ 
look random in the basis of eigenstates of
$U$) the fidelity operator $M_t$ becomes a matrix of independent gaussian random variables.
Then the average of the
product of four matrix elements (\ref{eq:FPSA}) can be estimated by a pair-contraction rule with
$\ave{\bra{p,\alpha}M_t\ket{j,\nu}\bra{q,\beta}M_t^\dagger\ket{k,\mu}} = 
\frac{1}{\Ne\Ns}\delta_{pk}\delta_{\alpha\mu}\delta_{qj}\delta_{\beta\nu}$ as
\begin{equation}
F^* = \frac{\Ns + \Ne + 4 + {\cal O}(1/\Ns) + {\cal O}(1/\Ne)}{(\Ns + 1)(\Ne + 1)}.
\label{eq:Fstar}
\end{equation}

We have  defined a reasonable measure for decoherence for the echo process 
and have determined that purity decays faster for near integrable system 
than for a mixing one. This fact is quite surprising in view of previous 
results on decoherence, but not so much if we consider that it follows 
closely the results obtained for fidelity.
We will proceed to apply our findings to a kicked Ising spin chain.

\section{Kicked Ising chain}

We shall now consider application of purity fidelity decay in a class of model systems namely one-dimensional
spin $1/2$ chains. We consider a ring of $L$ spins described by Pauli 
operators $\sigma^\alpha_j$, $j\in\{0,1,\ldots L-1\}$,
$\alpha\in\{\x,\y,\z\}$, with periodic boundary conditions $\sigma^\alpha_{j+L}\equiv \sigma^\alpha_j$.
In particular we concentrate on the example of Kicked Ising (KI) model \cite{Prosen01} with the hamiltonian
\begin{equation}
H_{\rm KI}(t) = \sum_{j=0}^{L-1} \left\{
J_\z \sigma^\z_j\sigma^z_{j+1} + \delta_p(t)(h_\x \sigma^\x_j 
+ h_\z \sigma^\z_j)\right\}
\label{eq:ham}
\end{equation}
where $\delta_{\rm p}(t)=\sum_{m=-\infty}^\infty\delta(t-m p)$ is a periodic delta function, generating the Floquet-map
\begin{equation}
U=\exp\left(-iJ_\z\sum_j \sigma^\z_j \sigma^\z_{j+1}\right)\exp\left(-i\sum_j(h_\x\sigma^\x_j + h_\z\sigma^\z_j)\right),
\label{eq:FM}
\end{equation}
where we take units such that $p=\hbar=1$,
depending on a triple of independent parameters $(J_\z,h_\x,h_\z)$. KI is {\em completely integrable} 
for longitudinal ($h_\x=0$) and transverse ($h_\z=0$) fields \cite{ki}, 
and has finite parameter regions of ergodic and non-ergodic behaviors in the thermodynamic limit $L\to\infty$ 
for a tilted field (see fig.~\ref{fig:fcc}). The non-trivial integrability of a transverse kicking field, which somehow
inherits the solvable dynamics of its well-known autonomous version \cite{ti}, is quite 
remarkable since it was shown \cite{ki} that the Heisenberg dynamics can be calculated 
explicitly for observables which are bilinear in Fermi operators 
$c_j = (\sigma^\y_j-i\sigma^\z_j)\prod_{j'}^{j'<j}\sigma^\x_{j'}$ with time 
correlations $C(t)$ decaying to the non-ergodic stationary values $D$ as 
$|C(t)-D| \sim t^{-3/2}$.
In order to test our predictions of purity fidelity decay by explicit calculation and/or
numerical experiment we consider a line in 3d parameter space, with fixed
$J_\z=1,h_\x=1.4$ and varying $h_\z$ exhibiting all different types of dynamics: 
(a) {\em integrable} for $h_\z = 0$, (b) {\em intermediate} (non-integrable and non-ergodic)
for $h_\z = 0.4$, and (c) {\em ergodic and mixing} for $h_\z = 1.4$.
In all cases we fix the operator $A=M:=\sum_j \sigma_j^\x$ which 
generates the following parametric perturbation of KI model
with 
\begin{eqnarray}
h_\x &\to& h_\x + \frac{(h_\x^2 + h_\z^2 h \cot h)}{h^2}\delta + {\cal O}(\delta^2),\label{eq:hpert}\\
h_\z &\to& h_\z + \frac{h_\x h_\z (1 - h\cot h)}{h^2}\delta + {\cal O}(\delta^2), \quad
{\rm where}\quad h=\sqrt{h_\x^2+h_\z^2} \nonumber
\end{eqnarray}
and vary the size $L$ and the perturbation strength $\delta$. 

For the sake of mathematical simplicity of the model we do not want to introduce extra
degrees of freedom to simulate the environment. Instead we logically split the
spin-chain by assigning a subset $\Js$ of $L_\c$ spins to the central system and the
complement $\Je=\Z_L\setminus \Js$ of $L_\e=L-L_\c$ spins to the environment.   
Then we have $\Ns = 2^{L_\c}, \Ne = 2^{L_\e}, {\cal N} = 2^L$.
In our numerical experiments and explicit analytical calculations we shall consider three
special cases (in the first two cases we have to assume that $L$ is even):
\begin{itemize}
\item {\bf (A)} {\em Alternating subchains}, where every second spin is assigned to the
central system $\Js^\A = \{0,2,4\ldots,L-2\}$. 
This situation is supposed to be a good model of a situation where all central system's 
degrees of freedom are directly coupled to the environment.
\item {\bf (C)} {\em Connected subchains}, where a connected half of the chain is assigned to the
central system $\Js^\C = \{0,1,2\ldots,L/2-1\}$. Here the central system is coupled to the
environment just at two ending points, namely $j=0$ and $j=L/2-1$.
\item {\bf (S)} {\em Single spin}, where the central system consists of a single spin $\Js^\S = \{0\}$,
i.e. a two-level quantum system coupled to a correlated many-body environment.
\end{itemize}

Our explicit calculations will be performed for a random initial state, in other words only state averaged
purity fidelity will be computed.

\subsection{Exact calculation for the integrable case}

Here we show how to extend our analytical approach \cite{ki} to solve purity fidelity decay
in the integrable case $h_\z = 0$. We start by a finite size version of the formalism, 
however the final results will be most elegant and simple in the thermodynamic limit $L\to\infty$.
The main quantity which we want to calculate is a state-averaged reduced stiffness 
$\ave{\tilde{D}}_\psi$ (\ref{eq:redstif}). As a first step in our calculation we have 
to construct a time average operator $\bar{A}$ associated to the perturbing operator $A$.
This can be done, as described in \cite{ki}, by means of the invariant space of the {\em adjoint}
map $U^{\rm ad}$, which is a unitary operator defined on a linear space of (bounded) observables as 
$U^{\rm ad} A = U^\dagger A U$. The complete basis of this invariant space can be constructed in terms 
of two sequences of $2L$ operators, $U_n$, $V_n$, $n\in \Z_{2L}$, namely
\begin{eqnarray}
U_n &=& \sum_{j=0}^{L-1}
\left\{ \begin{array}{lll}
     &\sigma^\y_j(\sigma^\x_j)_{n-1}\sigma^\y_{j+n}, & n \ge 1, \\
-\!\!\!\!&\sigma^\x_j, & n = 0, \\
     &\sigma^\z_j(\sigma^\x_j)_{-n-1}\sigma^\z_{j-n}, & n \le -1,
\end{array}\right. \label{eq:repr} \\
V_n &=& \sum_{j=0}^{L-1}
\left\{ \begin{array}{lll}
     &\sigma^\y_j(\sigma^\x_j)_{n-1}\sigma^\z_{j+n}, & n \ge 1, \\
     &1, & n = 0,\\
-\!\!\!\!&\sigma^\z_j(\sigma^\x_j)_{-n-1}\sigma^\y_{j-n}, & n \le -1,
\end{array}\right. \nonumber
\end{eqnarray}
where $(\sigma^\x_j)_k:=\prod_{l=1}^k \sigma^\x_{j+l}$ for $k \ge 1$,
$(\sigma^\x_j)_0:=1$, satisfying a Lie Algebra
\begin{eqnarray}
\left[ U_{m},U_{n} \right] &=& 2i ( V_{m-n} - V_{n-m} ), \nonumber\\
\left[ V_{m},V_{n} \right] &=& 0, \label{eq:comrel}\\
\left[ U_{m},V_{n} \right] &=& 2i ( U_{m+n} - U_{m-n} ). \nonumber
\end{eqnarray} 
We note the orthogonality 
\begin{equation}
\ave{U_n U_m} = \ave{V_n V_m} = L\delta_{nm},\quad \ave{U_n V_m} = 0.
\label{eq:orthUV}
\end{equation}
Now, the Floquet-map (\ref{eq:FM}) can be written in terms of the elements of the
algebra (\ref{eq:repr}) only, namely $U=\exp(-i J U_1)\exp( i h_\x U_0)$, and 
the complete basis of an invariant space
\begin{equation}
U^{\rm ad}\Psi(\varphi) = \Psi(\varphi)
\end{equation} 
can be, for arbitrary finite $L$, written as $\Psi(\varphi_l)$ with $\varphi_l = \frac{\pi l}{L}$, $l\in\Z_{2L}$,
and
\begin{eqnarray}
\Psi(\varphi_l) &=& \sum_{n=0}^{2L-1}(\ve{u}(\varphi_l)e^{in\varphi_l} + 
\ve{u}(-\varphi_l)e^{-in\varphi_l})\cdot
\ve{E}_n, \label{eq:Psi} \\
\ve{E}_n &\equiv& (U_n,U_{-n},2^{-1/2}(V_n - V_{-n})), \nonumber \\
\ve{u}(\varphi) &\equiv& (\cot\alpha - \cot\beta e^{-i\varphi},\cot\alpha - \cot\beta e^{i\varphi},\sqrt{2}i\sin\varphi),\nonumber
\end{eqnarray}
where we introduced the angles $\alpha \equiv 2J,\beta \equiv 2 h_x$. 
Note that $\Psi(\varphi_{2L-l})\equiv \Psi(\varphi_l)$ and the basis (of non-equivalent vectors 
$l=0,1,\ldots,L$) is orthogonal
\begin{equation}
\ave{\Psi^\dagger(\varphi_k) \Psi(\varphi_l)} = 2 L^2 
|\ve{u}(\varphi_k)|^2(\delta_{k,l} + \delta_{k,2L-l}).
\label{eq:orthpsi}
\end{equation}
The time average of an observable can be simply constructed by means of a projection
\begin{equation}
\bar{A} = \sum_{l=0}^{L} \frac{\ave{\Psi^\dagger(\varphi_l)A}}{\ave{\Psi^\dagger(\varphi_l)\Psi(\varphi_l)}}
\Psi(\varphi_l),
\label{eq:ave1}
\end{equation}
namely for the magnetization $A=M=U_0$ we have
\begin{equation}
\bar{M} = \frac{1}{L}\sum_{l=0}^{2L-1} \frac{{\rm Re\,}u_1(\varphi_l)}{|\ve{u}(\varphi_l)|^2}
\Psi(\varphi_l).
\label{eq:ave2}
\end{equation}
In the limit $L\to\infty$ the phase variable becomes continuous $\varphi\in[0,2\pi)$ and the 
sums (\ref{eq:ave1},\ref{eq:ave2}) go over to integrals over $\varphi$.
The first term of (\ref{eq:redstif}) is the stiffness $D = \ave{\bar{M}^2}$ 
which can be calculated explicitly in the limit $L\to \infty$ \cite{ki}. 
Finite size corrections are proven to be smaller than any power in $1/L$, i.e. they 
have to be {\em exponentially small}, since the argument of the sum which is
approximated by an integral over $\varphi$ is an $L$-independent {\em analytic} function of $e^{i\varphi}$
\begin{equation}
D = L \frac{{\rm max}\{|\cos\alpha|,|\cos\beta|\}-\cos^2\beta}{\sin^2\beta} + {\cal O}\left(\exp(-{\rm const}\, L)\right).
\label{eq:D}
\end{equation}
In order to calculate the remaining terms of the reduced stiffness (\ref{eq:redstif})
we have to compute the {\em reduced inner product} of a pair of operators, namely 
\begin{equation}
\ave{\ave{A}_\J \ave{B}_\J}_{\J'} = 2^{-|\J'|-2|\J|}\tr_{\J'}(\tr_\J A \tr_\J B)
\end{equation} 
where $\J$ is some subset of $\Z_L$ (either $\Js$ or $\Je$) and
$\J' = \Z_L \setminus \J$.
We start with the basis operators $U_n,V_n$, for which we find by direct inspection of the
structure (\ref{eq:repr}), that
\begin{eqnarray}
\ave{\ave{U_m}_\J \ave{U_n}_\J}_{\J'} 
&=& \ave{\ave{V_m}_\J \ave{V_n}_\J}_{\J'} = 
L f_n \delta_{mn}, \\
\ave{\ave{U_m}_\J \ave{V_n}_\J}_{\J'} &=& 0.\nonumber
\end{eqnarray}
Here the {\em structure function} $f_n$ is defined by the number of {\em sequential} sets 
${\cal C}(l_1,l_2)=\{l_1,l_1+1,\ldots,l_2\}$ in $\Z_L$ 
of length $|n|+1$ which do not intersect the set $\J$, divided by $L$
\begin{equation}
f_n = \frac{1}{L}|\{j\in \Z_L;{\cal C}(j,j+n)\cap \J=\emptyset\}|,
\end{equation}
and is extended to $\Z_{2L}$ by putting $f_{-n} \equiv f_{2L-n} := f_n.$
Using a {\em form factor} of sorts
\begin{equation}
F(\varphi) = \sum_{n=0}^{2L-1} f_n e^{i n\varphi},
\end{equation}
we can directly write the reduced inner products within the invariant family 
(\ref{eq:Psi}) generalizing (\ref{eq:orthpsi}), namely
\begin{equation}
\ave{\ave{\Psi(\varphi')^\dagger}_\J \ave{\Psi(\varphi)}_\J}_{\J'} 
= 2 L \ve{u}(\varphi')^* \cdot \ve{u}(\varphi)
\left(F(\varphi-\varphi')+F(\varphi+\varphi')\right).
\label{eq:rip}
\end{equation}
The calculation of (\ref{eq:redstif}), the reduced inner 
product of a time-averaged observable $\bar{A}$ with itself,
$\ave{\ave{\bar{A}}_\J^2}_{\J'}$ is of central importance. It can be computed straightforwardly 
using the expansion (\ref{eq:ave1}) and reduced inner products (\ref{eq:rip}). 
For the magnetization $M$ we obtain
\begin{eqnarray}
G_F:=\frac{1}{L}\ave{\ave{\bar{M}}_\J^2}_{\J'} 
=\frac{1}{2L^2}\sum_{k=0}^{2L-1}\sum_{l=0}^{2L-1}&&
\frac{{\rm Re} u_1(\varphi_k) {\rm Re} u_1(\varphi_l)
\ve{u}(\varphi_k)^*\!\cdot\!\ve{u}(\varphi_l)}{|\ve{u}(\varphi_k)|^2 |\ve{u}(\varphi_l)|^2}\times
\nonumber\\
&&\times\left(F(\varphi_l-\varphi_k)+F(\varphi_l+\varphi_k)\right).
\label{eq:GF}
\end{eqnarray}
The factor $L$ has been taken out in the definition of the coefficient $G_F$ in order to make it 
conveniently size-independent for the thermodynamic limit. In order to proceed with explicit 
calculations we have to derive the form factors for the
three different cases ({\bf A,C,S}). For each of them we have to compute two generally different reduced inner 
products with the form factors, $F_\c$ for $\J=\J_\c,\J'=\J_\e$, and $F_\e$ for $\J=\J_\e,\J'=\J_c$, 
corresponding respectively to the last two terms of (\ref{eq:redstif}).
\begin{itemize}
\item {\bf (A)}: Here, due to symmetry both structure functions are identical 
$f^\A_n = \frac{1}{2}\delta_{n,0}$ giving $F^\A_\c (\varphi) = F^\A_\e (\varphi) = \frac{1}{2}$.
\item {\bf (C)}: Again, we have a symmetry between the central system and the environment, so
$f^\C_n = {\rm max}\{\frac{1}{2}-\frac{|n|}{L},0\}$ and 
$F^\C_\c(\varphi) = F^\C_\e(\varphi) = \pi \delta_{L/2}(\varphi)$ where $\delta_m(\varphi)$ 
is a fat periodic delta function
\begin{equation}
\delta_m(\varphi) = \frac{1}{2\pi m}\left(\frac{\sin(m\varphi/2)}{\sin(\varphi/2)}\right)^2,
\end{equation} 
with the limit $\lim_{m\to\infty}\delta_m(\varphi) = 
\sum_{k=-\infty}^\infty \delta(\varphi - 2\pi k)$.
\item {\bf (S)}: Here, we need to consider two different structure functions
$f^\S_{\c,n}=(L-1-|n|)/L, f^\S_{\e,n}=\frac{1}{L}\delta_{n,0}$ giving for the form factors 
$F^\S_\c(\varphi) = \frac{1}{L}$,
$F^\S_\e(\varphi) = \frac{2\pi(L-1)}{L}\delta_{L-1}(\varphi)$.
\end{itemize}
Due to linear dependence of $G_F$ on $F(\varphi)$
(\ref{eq:GF}) we need to compute only two types of the reduced inner products 
besides the non-reduced one (\ref{eq:D}), namely for a constant form factor
$F(\varphi) = 1$ and for a fat-delta form factor $F(\varphi) = \delta_m(\varphi)$.
Tedious but straightforward calculation gives (within accuracy ${\cal O}(\exp(-{\rm const}\,L))$ beyond {\em all orders})
\begin{eqnarray}
G_1 &=& \left\{
\begin{array}{ll}
2\cot\beta\sin^2(\alpha/2); & \cos\alpha > \cos\beta, \cos\alpha + \cos\beta > 0,\\
-\tan(\beta/2); &             \cos\alpha < \cos\beta, \cos\alpha + \cos\beta > 0,\\
\cot(\beta/2); &              \cos\alpha > \cos\beta, \cos\alpha + \cos\beta < 0,\\
2\cot\beta\cos^2(\alpha/2); & \cos\alpha < \cos\beta, \cos\alpha + \cos\beta < 0,
\end{array}\right. \nonumber \\
G_{\delta_m} &=& G_{\delta} - \frac{1}{4\pi m}, \quad G_{\delta} := \frac{D}{2\pi L}.
\end{eqnarray}
The state averaged purity fidelity in the regime of linear response 
(\ref{eq:lra}) can be therefore written as
\begin{eqnarray}
&&\ave{F_\P(t)}_\psi = 1 - \frac{2\delta^2 \Ns \Ne L}{(\Ns+1)(\Ne+1)} 
\tilde{G} t^2 + {\cal O}(\delta^4),\label{eq:anal}\\
&&\tilde{G} = \frac{1}{L}\left(\ave{\bar{M}^2} - 
\ave{\ave{\bar{M}}_{\J}^2}_{\J'} - \ave{\ave{\bar{M}}_{\J'}^2}_{\J}\right) \nonumber
\end{eqnarray}
where the coefficient in the three different cases of interest reads
\begin{eqnarray}
\tilde{G}^\A &=& 2\pi G_{\delta} - G_1,\nonumber \\
\tilde{G}^\C &=& \frac{1}{L},\label{eq:Gtil}\\
\tilde{G}^\S &=& \frac{1}{L}\left(2\pi G_{\delta} - G_1 + \frac{1}{2}\right).\nonumber
\end{eqnarray}
It is very interesting to note that only in case {\bf (A)} where all environmental degrees of 
freedom are coupled to the central system and vice versa, the resulting expression for 
purity decay coefficient $\tilde{G}$ is non-vanishing in the
thermodynamic limit $L\to\infty$, so the purity-fidelity decays qualitatively in the same way 
as the (non-reduced) fidelity \cite{Prosen01}, namely the exponent of
quadratic decay in time has the same $L$ dependence.
In the other two cases {\bf (C,S)} where the interaction between the 
central system and the environment is {\em local} one has a subtle cancellation of leading 
order terms giving the resulting coefficient which vanishes as $\propto 1/L$ in the thermodynamic limit thus making
the purity-fidelity to decay still quadratically but on much longer time scale (by a factor of $L$) as compared to the
time scale of fidelity decay \cite{Prosen01}.
Of course, our analytical results for the integrable case only give us a time-scale of the purity-fidelity decay
\begin{equation}
\tilde{\tau}_{\rm ne} = \frac{1}{\delta}\sqrt{\frac{(\Ns+1)(\Ne+1)}{\Ns \Ne 
L \tilde{G}}},
\label{eq:taune}
\end{equation}
but they cannot tell us anything about the global behaviour of $\ave{F_\P(t)}_\psi$ beyond the regime of linear response.
These analytical results are clearly confirmed by direct numerical simulations reported in next subsection.

\subsection{Numerical calculations in the general case}

In the generally non-integrable cases of kicked Ising model we calculate the {\em partial correlation sums}
\begin{equation}
S_\J(t) = \frac{1}{L}\ave{\ave{\Sigma A_t}_\J \ave{\Sigma A_t}_\J}_{\J'}
\end{equation} 
and the {\em total purity correlator}
\begin{equation}
S_\P(t) = \frac{1}{L}\sum_{t',t''=0}^{t-1} C_\P(t',t'') = S_{\emptyset}(t) + S_{\Z_L}(t) - S_{\J}(t) - S_{\J'}(t)
\end{equation}
by means of numerical simulation. With these we can calculate purity 
fidelity and  compare to the directly numerically simulated
purity fidelity decay (and to the analytical calculation in the integrable case).
Note that $S_{\emptyset}(t)=\frac{1}{L}\ave{[\Sigma A_t]^2}=\frac{1}{L}\sum_{t',t''=0}^{t-1}\ave{A A_{t'-t''}}$ is 
the usual (non-reduced) integrated correlation function and $S_{\Z_L}(t)\equiv 0$ due to $\ave{A_t}=\ave{A}=0$.
We work with a random initial states, hence in the linear response regime (sufficiently small
$\delta$) we have an identity (\ref{eq:lra}) connecting the state-averaged purity fidelity decay to the purity correlator.
Asymptotic ($\delta\to 0$) exactness of this relation has been carefully checked in all different cases of interest since
it provided a crucial test of our numerical procedures.

In the first set of numerical calculations we check the decay of correlations, either with respect to 
the full trace (non-reduced) inner product or with respect to the reduced inner product, while in the second set 
of calculations we explicitly compute the purity fidelity decay $\ave{F_\P(t)}_\psi$ and compare the exponents with 
our predictions based on correlation decay. Calculations are performed with three different system sizes namely 
$L=12,16,20$. Since we want the perturbation strength to have certain size $L$-independent effects
we scale it by fixing $\delta'=\delta\sqrt{L/L_0}$ when varying $L$ where we choose $L_0:=24$. 
For example, in such a case the predicted exponents of the fidelity decay \cite{Prosen01} do not depend on $L$.

\begin{figure}
\hbox{\hspace{1in}
\includegraphics[width=7.5in]{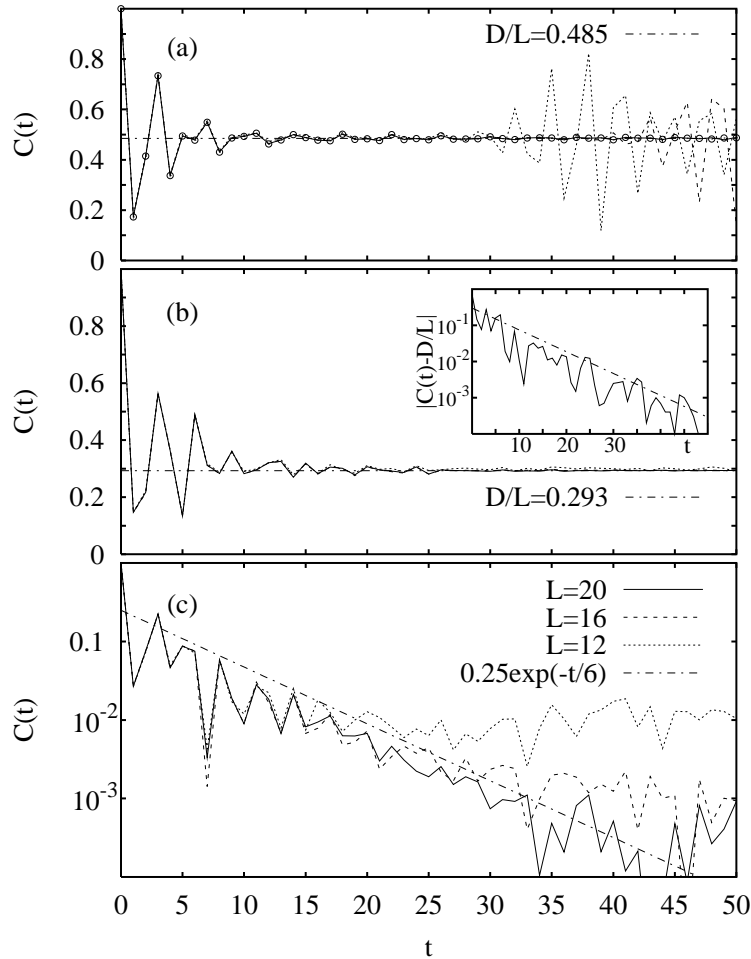}}
\caption{Correlation decay for three cases of KI: (a) integrable $h_z=0$, (b) intermediate $h_z=0.4$, and (c)
ergodic $h_z=1.4$, for different sizes $L=20,16,12$ (solid-dotted connected curves, 
almost indistinguishable in (a,b)). Circles (a)
show exact $L=\infty$ result. Chain lines are theoretical/suggested asymptotics (see text). 
}
\label{fig:fcc}
\end{figure}

In fig.~\ref{fig:fcc} we examine the decay of non-reduced correlations of the magnetization $C(t) = \ave{M M_t}/L$ for the
three cases: (a) integrable $h_z=0.0$, (b) intermediate $h_z=0.4$, and (c) ergodic and mixing $h_z=1.4$.
We find that $C(t)$ has a non-vanishing plateau (stiffness $\bar{C}=D/L$) in the integrable and intermediate
non-integrable case, where the correlation function in integrable 
case agrees excellently with the analytical result for $L\to\infty$ \cite{ki}, 
while in ergodic and mixing case we find exponential decay of correlations. We note an interesting distinction between
correlation decays in integrable and intermediate cases, namely in the integrable case (a) the relaxation of $C(t)$
towards the plateau value $D/L$ is a power law $\sim t^{-3/2}$ whereas in the intermediate case (b) it 
looks like an exponential (see inset of fig.~\ref{fig:fcc}b). 

\begin{figure}
\centerline{\includegraphics[width=4.5in]{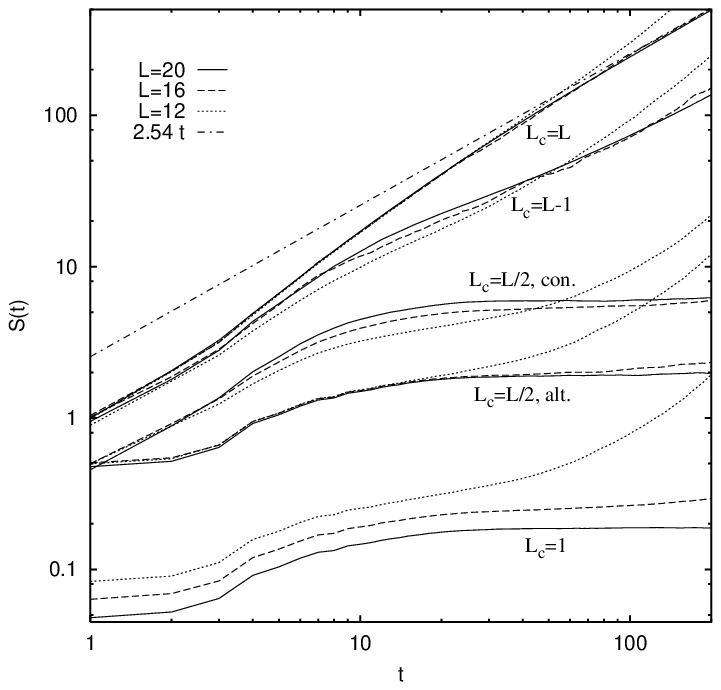}}
\caption{Partial correlation sums in the mixing regime $h_z=1.4$ for different divisions $\Z_L =\J \cup \J'$.
The structure (alternating {\bf A} vs. continuous {\bf C}) and/or the number $L_\c$ of elements of the set 
$\J'$ is  indicated in the label near each triple of curves 
(full for $L=20$, dashed for $L=16$ and dotted for $L=12$).
}
\label{fig:fce}
\end{figure}

\begin{figure}
\hbox{\hspace{1in}
\includegraphics[width=4.5in]{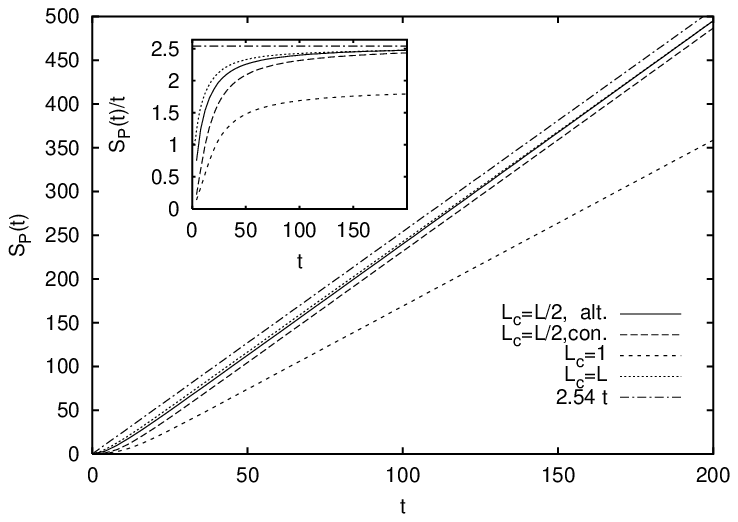}}
\caption{Purity correlator $S_\P(t)$ in the mixing regime $h_z=1.4$ for different cases ({\bf A,C,S}, 
and total (non-reduced) correlator, for full, long-dashed, short-dashed, and dotted curves, respectively) 
as compared to the asymptotic linear increase (chain line). 
}
\label{fig:fse}
\end{figure}

In fig.~\ref{fig:fce} we show partial correlation sums $S_\J(t)$ for the ergodic and mixing case (c) for different 
reducing sets $\J$ appearing in the three cases ({\bf A,C,S}) of divisions.
We note that $S_\J(t)$ increases linearly in $t$ only for the non-reduced case $\J=\emptyset$ where (non-reduced)
correlation function is homogeneous in time, whereas in all other cases $S_\J(t)$ increases slower than linear, 
in fact in most cases it quickly saturates to a maximum value which does not increase with increasing size $L$. 
Therefore, for large system sizes $L$ in the ergodic and mixing situation the purity correlator becomes 
determined by the total (non-reduced) correlator only [confirming eq. (\ref{eq:eqsig})], namely
$S_\P(t) = s t$, where $s = 2.54$. This is illustrated in fig.~\ref{fig:fse}.

\begin{figure}
\centerline{\includegraphics[width=4.5in]{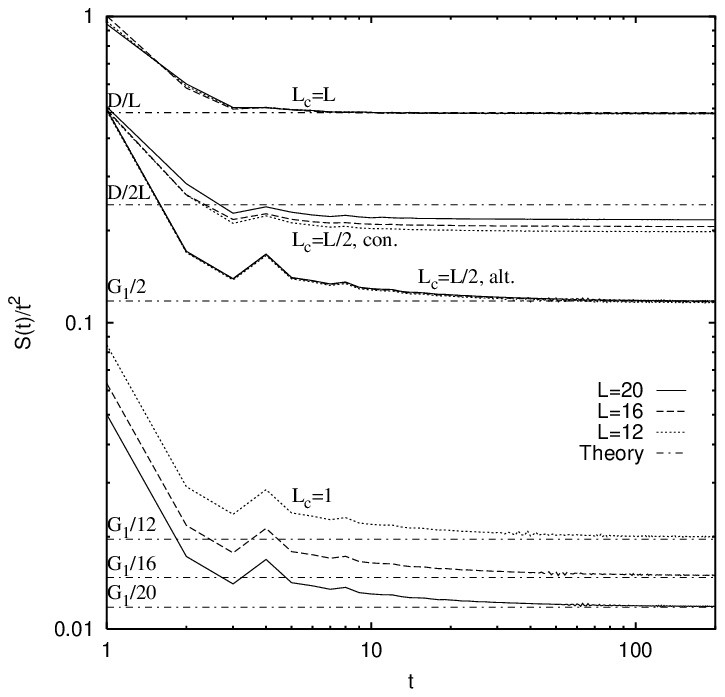}}
\caption{Partial correlation sums in the integrable regime $h_z=0.0$ compared to the analytically computed coefficients for
different divisions indicated by labels describing the structure (con./alt.) and number of elements of the set $\J'$.
}
\label{fig:fci}
\end{figure}

\begin{figure}
\hbox{\hspace{1in}
\includegraphics[width=4.5in]{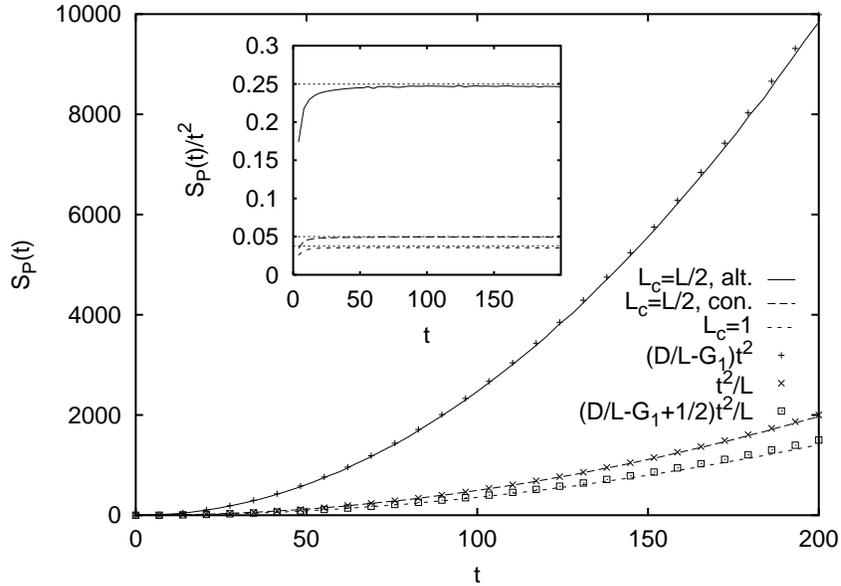}}
\caption{Purity correlator $S_\P(t)$ 
in the integrable regime $h_z=0.0$ compared with an analytical expressions (sampled symbols) for
different cases of divisions ({\bf A,C,S}). In the inset we emphasize the quadratic increase by plotting
$S_\P(t)/t^2$ and comparing to theoretical coefficients (\ref{eq:Gtil}) (dotted).
}
\label{fig:fsi}
\end{figure}

\begin{figure}
\hbox{\hspace{1in}
\includegraphics[width=4.5in]{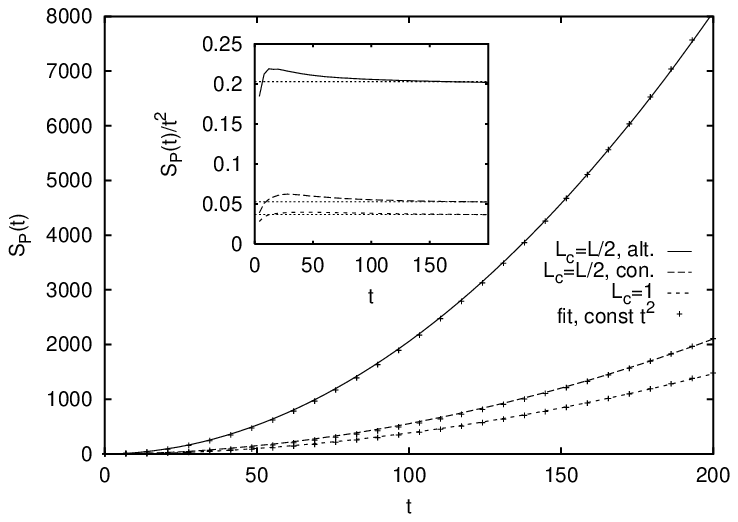}}
\caption{Same as in figure (\ref{fig:fsi}) but for the intermediate case (b).
Since we have no analytical predictions here, we compare the data by best 
fitting quadratic functions (sampled symbols, and dotted in the inset) 
whose coefficients are later used for comparison to purity fidelity decay. 
}
\label{fig:fsn}
\end{figure}

In figs.~\ref{fig:fci},\ref{fig:fsi} we show analogous results on partial correlation sums and purity correlators in
the integrable case (a). Here, $S_\P(t)\propto t^2$. Numerical results are compared with analytical expressions 
(\ref{eq:anal},\ref{eq:Gtil}) and the agreement is very good. In 
fig.~\ref{fig:fsn} we also show the purity correlators 
in the intermediate regime (b) where the results are qualitatively very similar to the integrable case (a), 
namely we have the quadratic growth of purity correlators $S_\P(t) \propto t^2$.

In the second set of numerical experiments we calculate the purity fidelity $F_\P(t)$ with respect to a random 
initial state $\ket{\psi}$ and average the result over a sufficient number of realizations of random initial state
such that statistical fluctuations are negligible. We note that $F_\P(t)$ is {\em self-averaging}, namely by increasing
the dimension of the Hilbert space i.e. increasing the size $L$ 
the purity fidelity of a single random initial state converges to the state average $\ave{F_P(t)}_\psi$.
Here we are interested not only in the linear response regime but also in the global behaviour of purity fidelity,
so we chose to display the results for the (scaled) perturbation strength $\delta'=0.01$.

\begin{figure}
\centerline{\includegraphics[width=4.5in]{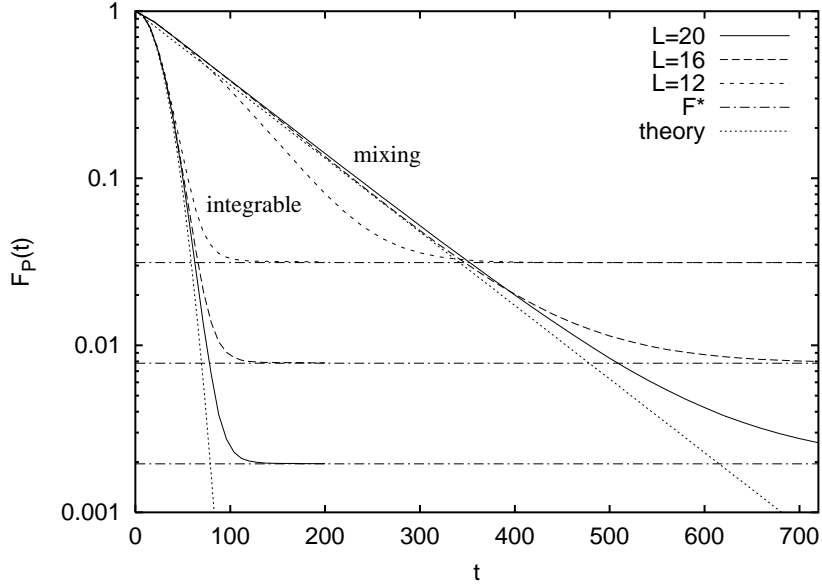}}
\caption{Purity fidelity decay: comparision between mixing $h_z=1.4$ and integrable regime $h_z=0$, at scaled perturbation
$\delta'=0.01$. Sampled symbols give the predicted exponential (\ref{eq:FPem}) in the mixing regime and a suggested gaussian with a theoretically computed time-scale (\ref{eq:taune}) in the integrable regime, while horizontal chained lines
give the plateau values (\ref{eq:Fstar}).
}
\label{fig:fpf}
\end{figure}

\begin{figure}
\centerline{\includegraphics[width=4.5in]{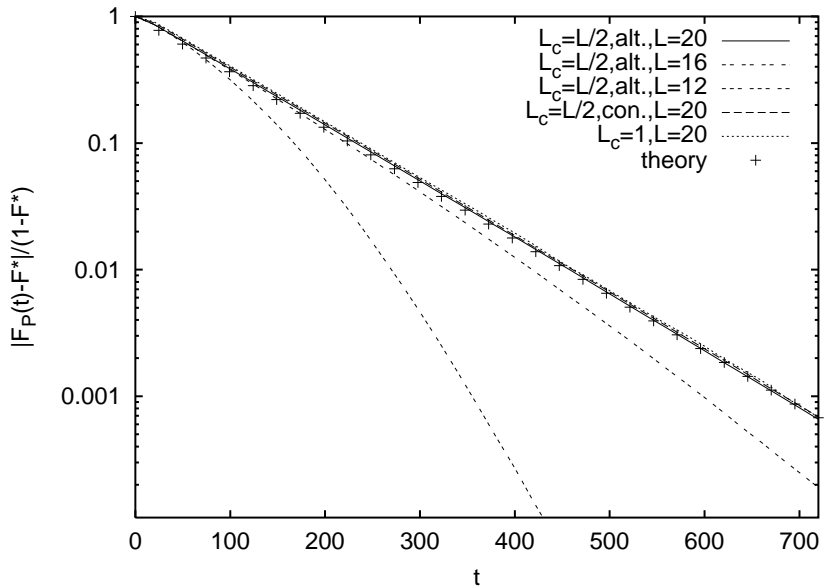}}
\caption{Purity fidelity decay in mixing regime $h_z=1.4$ for different types of division and different sizes $L$ 
(indicated in a legend) and $\delta'=0.01$. Theoretical decay (for $L\to\infty$) is given by sampled symbols.}
\label{fig:fpe}
\end{figure}

\begin{figure}
\centerline{\includegraphics[width=4.5in]{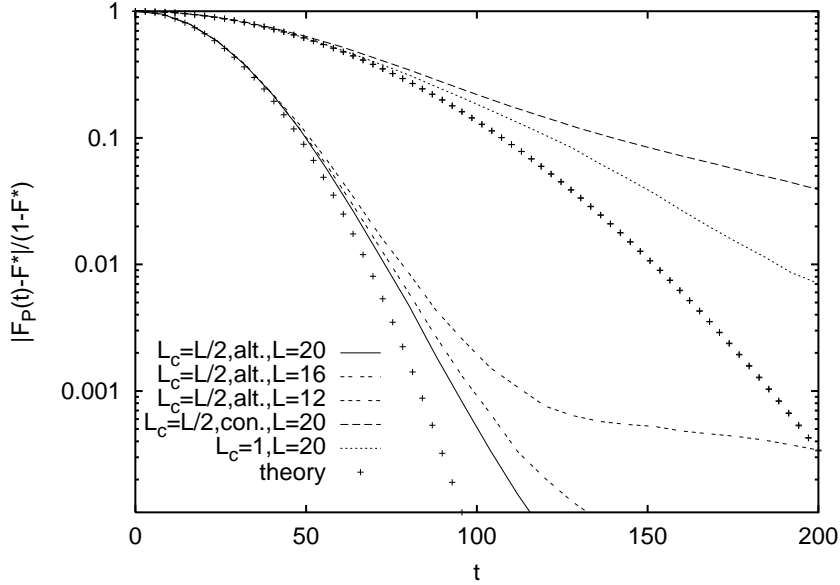}}
\caption{Same as in fig.~\ref{fig:fpe} but for integrable regime (a). Note that theoretical (gaussian extrapolated) curves for cases ({\bf C,S}) are 
practically indistinguishable.}
\label{fig:fpi}
\end{figure}

\begin{figure} 
\centerline{\includegraphics[width=4.5in]{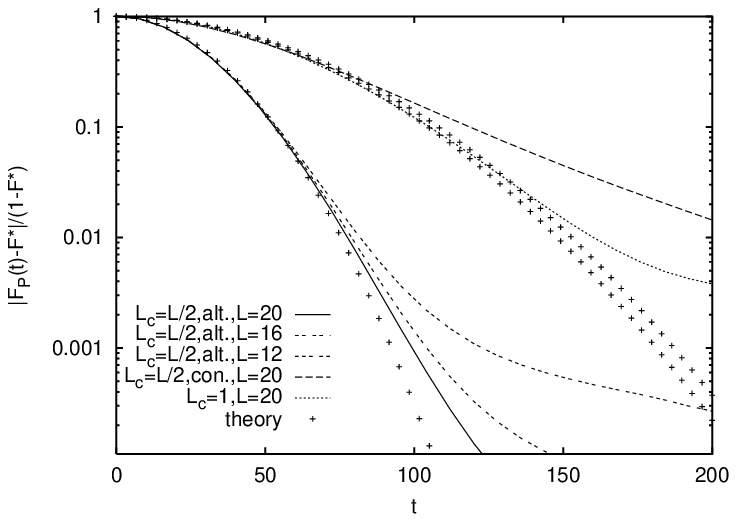}} 
\caption{Same as in fig.~\ref{fig:fpe} but for intermediate regime (b).}
\label{fig:fpn} 
\end{figure}

According to our relation (\ref{eq:lra}) and the behaviour of purity correlators we predict (and find!) that purity 
fidelity decay is faster in integrable and intermediate cases (a,b) than in the ergodic and mixing case (c). 
This result is most clear-cut in the case of division ({\bf A}). See fig.~\ref{fig:fpf} for a comparison of purity
fidelity decay in integrable and mixing cases. We have found very good agreement with the predicted exponential decay
(\ref{eq:FPem}) in the mixing case (c), whereas in the integrable regime (a) numerical results suggest a global gaussian 
decay of purity fidelity (similarly to the fidelity decay \cite{Prosen01}) with an analytically computed exponent 
(\ref{eq:taune}).
Due to finite Hilbert space dimension we find a saturation of purity fidelity for very long times at the
plateau value which has been computed in eq. (\ref{eq:Fstar}). In order to avoid the trivial effect of a constant
term $F^*$ in (\ref{eq:FPSA}) we subtract it and in the following plots show the quantity $|F_\P(t)-F^*|/(1-F^*)$.
In fig.~\ref{fig:fpe} we show that purity fidelity decay in ergodic and mixing regime (c) is independent of the type
of division ({\bf A,C,S}). In figs.~\ref{fig:fpi},\ref{fig:fpn} we show an analogous comparison of purity fidelity decay 
in integrable and intermediate cases (a,b) for different divisions and show that, in all cases, short time behaviour 
is well reproduced by the linear response coefficients given by purity correlators. For longer times a global 
gaussian behaviour with theoretical exponents work quite well (in particular for case ({\bf A})).

\section{Discussion}

We have analyzed the properties of decoherence in the framework of the 
question of reversibility with a perturbed time reversed Hamiltonian.
Up to now only the state correlation function, usually called fidelity,
has been discussed in this context. The calculations were carried out in the framework of 
partial tracing over the environment and unitary time-evolution.
The remaining density operator on the subspace was analyzed in terms 
of the trace of its square, usually called purity. It is more
convenient to use this quantity rather than entropy itself, because its 
analytic form allows to perform explicit calculations to a much larger extent
than would be the case for the entropy, where the logarithm complicates things.

We apply the techniques developed for the calculation of fidelity to a Hilbert space 
which is a product space. Keeping track of the respective indices we can
readily perform the partial traces needed, and find that they relate to slightly
modified correlation functions of quantum observables. 
This allows to compare the results for purity-fidelity with those of fidelity \cite{Prosen01}. 

As for the fidelity it becomes apparent that the decay time of correlations is 
a relevant short time scale, and this induces for mixing systems an exponential decay
with a time scale that is mainly related to the strength of the perturbation.
This implies a linear decay after the end of the trivial short time `perturbative' regime 
which leads to a quadratic decay always. For integrable systems on the other hand correlations
do not decay and thus the quadratic decay of purity-fidelity survives
beyond the perturbation regime. This amounts to the central result that
decoherence will be {\em faster} for an integrable system, than for a chaotic 
one in strict analogy to the findings for fidelity.

The techniques developed are applied to kicked spin chains where analytic results
for the integrable case and numerical results for the correlation functions of observables
can easily be obtained. In the thermodynamic limit we obtain exponential decay
of purity-fidelity for the mixing case and gaussian decay for the for the 
integrable case with uniform coupling to the environment, which is simulated
by assigning alternatively one spin to the environment and one to the central system.
If the spin chain is simply cut in two or if a single spin is associated to the central
system we find deviations from the gaussian shape, but the decay rate associated with
the quadratic behaviour at short but non-perturbational times is correctly reproduced
thus maintaining the fact that even in these situations coherence decays faster in
the integrable case than in the mixing one. 

The central message is thus that decoherence in echo situations follows fidelity. Indeed the 
reversibility of decoherence under perturbation of the reversed Hamiltonian 
is better for mixing systems than for integrable ones, and the time 
scales on which this occurs depend sensitively on the strength of this perturbation. 

\section*{Acknowledgements}

TP acknowledges financial support by the Ministry of Education, Science and Sport of Slovenia. THS thanks Physics Department, Unversity of Ljubljana for hospitality, and acknowledges financial support from the grants CONACyT 25192-E and
DGAPA IN-102597.

\end{document}